\documentclass[a4paper, 12pt]{article}
\usepackage{setspace}
\usepackage{amsmath}
\usepackage{graphicx}
\usepackage{amsfonts}
\usepackage{amssymb}
\usepackage{graphicx}
\usepackage{epstopdf}
\usepackage{epsfig}
\usepackage{amsmath}
\usepackage{amssymb}
\usepackage{amsthm}
\usepackage{mathptmx}
\usepackage{multirow}
\usepackage{cite}
\usepackage{xcolor}
\usepackage{float}

\numberwithin{equation}{section}

\theoremstyle{plain}
\newtheorem{thm}{Theorem}[section]
\newtheorem{cor}[thm]{Corollary}

\newtheorem{rem}[thm]{Remark}

\title{\bf Multiple soliton solutions and similarity reduction of a (2+1)-dimensional variable-coefficient Korteweg--de Vries system}
\author{Yaqing Liu$^1$\footnote{Email: liuyaqing@bistu.edu.cn.}, ~~
Linyu Peng$^2$\footnote{Email: l.peng@mech.keio.ac.jp, corresponding author.} \\
{\small\it $1.$ School of Applied Science, Beijing Information Science and Technology University,} \\
{\small\it  Beijing 100192, China}\\
{\small\it $2.$ Department of Mechanical Engineering, Keio University, Yokohama 223-8522, Japan}}

\begin{document}

\maketitle


\begin{abstract} In this paper, we study the novel nonlinear wave structures of a (2+1)-dimensional variable-coefficient Korteweg--de Vries (KdV) system by its analytic solutions. Its $N$-soliton solution are obtained via Hirota's bilinear method, and in particular, the hybrid solution of lump, breather and line soliton are derived by the long wave limit method. In addition to soliton solutions, similarity reduction, including similarity solutions (also known as group-invariant solutions) and non-autonomous third-order Painlev\'e equations, is achieved through symmetry analysis.  The analytic results, together with illustrative wave interactions, show interesting physical features, that may shed some light on the study of  other variable-coefficient nonlinear systems.
\end{abstract}


\section{Introduction}

Nonlinear partial differential equations play a crucial role in modeling wave phenomena that arise in various fields, such as fluid mechanics, nonlinear optics, plasma physics, condensed matter physics, etc. Methods for constructing their analytical solutions, that can explicitly describe the dynamical behavior, have attracted much attention, among which there are the Hirota's bilinear method \cite{Hirota2004,Kodama2017}, Darboux transformation \cite{GLL2012,XS2017},  the inverse scattering transformation \cite{AKNS1974,TD2013}, the multiple exp-function method \cite{LZH2018}, dressing method \cite{MPW2016, KZ2017},  symmetry method \cite{JM1983,Olver1993, Hydon2000,BA2002, Peng2017}, and so on. In the current paper, we will mainly be focused on soliton solutions by Hirota's bilinear method and similarity solution by symmetry reduction.

Hirota's bilinear method has been investigated by many scholars (see, e.g., \cite{FMZ2022, RMH2022}). For integrable nonlinear evolution equations,  Hirota's bilinear method can be used to construct  $N$-soliton solutions applying the superposition of solutions, and has been extended to obtain breather, lump and their interaction solutions \cite{LIMMAK2022},
higher-order localized wave \cite{FCH2022},  rational and semi-rational solution \cite{HG2022}. It was also employed for searching the localized waves of nonlocal evolution equations \cite{YF2020, ZL2022} and variable-coefficient evolution equations \cite{KM2021, Osman2019,WS2022}.

On the other hand, continuous symmetries have also been widely applied to the analysis of differential equations (see, e.g., \cite{BC1969,Olver1977,OR1987}). They can lead to exact solutions or reductions of differential equations \cite{Olver1993,BA2002}, and are closely relevant to their integrability  (see, e.g., \cite{MSS1991}).    
In recent years, symmetry analysis of  variable-coefficient evolution equations has attracted much attention.
For instance, in \cite{AS2021, AMS2022}, Mohamed and co-authors investigated lump soliton, solitary waves and exponential solutions of the (3+1) dimensional variable-coefficients Kudryashov--Sinelshchikov  equation and the (2+1)-dimensional variable-coefficient Bogoyavlensky--Konopelchenko equation  by similarity reduction using their symmetries.
In \cite{WLXZ2018}, a variable-coefficient Davey--Stewartson  system was studied, where
the optimal system of symmetries was obtained with adjoint representation. Variable-coefficient Davey--Stewartson  system that admits the Kac-Moody-Virasoro symmetry was proposed in \cite{GO2016}.
In \cite{XYX2020},  nonlocal symmetries of the coupled variable-coefficient Newell--Whitehead system were used to calculate its group-invariant solutions.
 However,
few studies have been conduced on the dynamics of higher-order localized waves. In the current paper, we focus on nonlinear wave structures of the following (2+1)-dimensional variable-coefficient Korteweg--de Vries (KdV)  system
\begin{equation}\label{eq:vckdv}
 \left\{
\begin{array}{l}
\rho(t)u_t+3\mu(t)(uv)_x+\sigma(t)u_{xxx}=0,\vspace{0.2cm}\\
u_x=v_y,
\end{array}
\right.
\end{equation}
and particularly  its integrable variant with both $\mu(t)$ and $\sigma(t)$ constant functions,
using both Hirota's bilinear method and symmetry analysis.
Here, $u$, $v$ are the dependent variables and $x$, $y$ and $t$ are the independent variables. The notations $u_t=\partial u/\partial t$,  $u_x=\partial u/\partial x$ and so forth are adopted in the current paper. The functions $\rho(t)$, $\mu(t)$ and $\sigma(t)$ are known but arbitrary  functions of $t$ which are assumed to be smooth enough; when $\rho(t)=\mu(t)=\sigma(t)= 1$, the system  \eqref{eq:vckdv} reduces to the constant-coefficient (2+1)-dimensional KdV system \cite{Wang2016,LMDL2016}, derived by Boiti {\it et al.}  using the weak Lax pair \cite{BLMP1986}. Furthermore,  the system  \eqref{eq:vckdv} reduces to the (1+1)-dimensional KdV equation by setting $v=u$ and $x=y$.



 In Section \ref{sec:Pai},  integrability condition of \eqref{eq:vckdv} is analyzed by Painlev\'{e} analysis and in what follows we will be focused on its integrable version with constant coefficients $\mu$ and $\sigma$. In Section \ref{sec:sol}, $N$-soliton solutions and hybrid interaction of line solitons, and breather and lump solitons are obtained through Hirota's bilinear method, while in Section \ref{sec:sym}, we invoke the symmetry method to derive Lie point symmetries and obtain the corresponding similarity solutions. In particular, a PDE (see Eq. \eqref{eq:newintpde}  or system \eqref{eq:redx5} in the potential form below) passing the Painlev\'e test is derived by symmetry reduction, that reads
 \begin{eqnarray}
U_r=\left(\frac{a(y)-\sigma U_{rr}}{3\mu U}\right)_y,
\end{eqnarray}
where $r,y$ are the independent variables and $U$ is the dependent variable, and $a(y)$ is an arbitrary function.  Further symmetry reduction shows that it can be reduced to non-autonomous third-order Painlev\'e equations, which are in the form of  Chazy's classification of third-order ODEs by Painlev\'e analysis but not included in Chazy's 13 classes. The final Section \ref{sec:con} is dedicated to conclusion.

\section{Painlev\'{e} analysis of the variable-coefficient KdV system}

\label{sec:Pai}



In this section, we study integrability condition of the (2+1)-dimensional variable-coefficient KdV system \eqref{eq:vckdv}  through Painlev\'e analysis (see, e.g., \cite{BKS1990, Wazwaz2019}).  Let $u=m_y$ and $v=m_x$, and the system \eqref{eq:vckdv} becomes
\begin{equation}\label{eq:4thpde}
\rho(t) m_{yt}+3\mu(t)(m_xm_y)_x+\sigma(t)m_{xxxy}=0,
\end{equation}
that is assumed to admit a solution as a Laurent expansion
about a singular manifold $\phi=\phi(x, y, t)$ as follows
\begin{equation}\label{eq:m}
m(x,y,t)=\phi^{-n}(x,y,t)\sum_{j=0}^{\infty}m_j(x,y,t){\phi^{j}(x,y,t)}, \quad n>0.
\end{equation}
Here, $n$ is determined by a leading-term analysis and balancing the dominant terms $(m_xm_y)_x$ and $m_{xxxy}$. Straightforward calculation gives $n=1$ and
\begin{equation}\label{eq:m0}
m_0=\frac{2\sigma(t)}{\mu(t)}\phi_x.
\end{equation}
Substituting $m=m_0 \phi^{-1}+q\phi^{r-1}$ back to Eq. \eqref{eq:m} and balancing the most singular term again, we obtain
\begin{equation} \label{eq:pa}
q\Big(3\mu(t)(4m_0(r-1)\phi_x^2\phi_y-2m_0(r-1)(r-2)\phi_x^2\phi_y) +\sigma(t)(r-1)(r-2)(r-3)(r-4)\phi_x^3\phi_y\Big)=0.
\end{equation}
Combining  Eqs. \eqref{eq:m0} and \eqref{eq:pa}, $q$ is arbitrary when $ r=-1, 1, 4 $ and $6$, which are the resonant points. Substitution of \eqref{eq:m} into Eq. \eqref{eq:4thpde} then amounts to the recursion relations for the $m_j$, which take the form of coupled partial differential equations.
Finally, we observe that explicit expressions for $m_2$, $m_3$ and $m_5$ and compatibility condition
to ensure integrability requires $\mu(t) = \mu$, $\sigma(t) = \sigma$ to be non-zero constants, but $\rho(t)$ remains to be an arbitrary 
function of $t$. Consequently, $m_1$, $m_4$ and $m_6$ are  arbitrary functions.

To summarize up, we conclude that the following special (2+1)-dimensional time-dependent variable-coefficient KdV system
\begin{equation}\label{eq:vckdv0}
 \left\{
\begin{array}{l}
\rho(t)u_t+3\mu(uv)_x+\sigma u_{xxx}=0,\vspace{0.2cm}\\
u_x=v_y,
\end{array}
\right.
\end{equation}
is integrable as passes the Painlev\'{e}  test, where $\mu$ and $\sigma$ are constants. In the rest of the paper, we will be focused on studies of analytic solutions of the integrable system
\eqref{eq:vckdv0}.


\section{Bilinear representation  and $N$-solitons}

\label{sec:sol}
The integrable (2+1)-dimensional time-dependent variable-coefficient KdV system
 \eqref{eq:vckdv0}  can be written in bilinear form
\begin{equation}
\Big(\rho(t)D_yD_t+\sigma D_x^3D_y+3\alpha \mu D_x^2+3\beta \mu D_xD_y\Big)f\cdot f=0
\end{equation}
by the transformation
\begin{equation} \label{eq:trans}
 \left\{
\begin{array}{l}
u=\alpha+\frac{2\sigma}\mu\big({\rm log}(f)\big)_{xy},\vspace{0.2cm}\\
v=\beta+\frac{2\sigma}\mu\big({\rm log}(f)\big)_{xx},\\
\end{array}
\right.
\end{equation}
where both $\alpha$ and $\beta$ are real-valued parameters, and 
 Hirota's bilinear differential operators are defined by
\begin{eqnarray}{\label{eq-2-3}}
D_x^m D_y^n D_t^s a\cdot b=(\partial_x -\partial_{x'})^m(\partial_y -\partial_{y'})^n(\partial_t -\partial_{t'})^s a(x,y,t)b(x',y',t')|_{x=x', y=y', t=t'}.
\end{eqnarray}
The function $f$ can have the general form as
\begin{eqnarray}{\label{e14}}
f_N=\sum_{\varsigma=0,1}\exp\left(\sum_{j=1}^N \varsigma_j \theta_j+\sum_{1\leq s<j}^N \varsigma_s \varsigma_j A_{sj} \right),\quad N=1,2,\ldots,
\end{eqnarray}
where 
\begin{eqnarray}{\label{e15}}
\theta_j=k_{{j}}x+p_{{j}}y-\omega_j(t)+\theta_{0j} \text{ with }   \omega_j(t)=\int \frac{\sigma k_j^3p_j +3\beta \mu k_j p_j+3\alpha \mu k_j^2}{p_j \rho(t)}dt
\end{eqnarray}
 for $j=1,2,\ldots, N$, and
\begin{eqnarray}{\label{e16}}
\exp(A_{sj})=\frac {  \sigma k_s k_j  (k_s - k_j)(p_s-p_j)p_s p_j - \alpha \mu(k_j p_s-k_s p_j)^2}{ \sigma k_s k_j  (k_s + k_j)(p_s+p_j)p_s p_j - \alpha \mu(k_j p_s-k_s p_j)^2}, ~~1 \leq s<j\leq N.
\end{eqnarray}
Here, $k_j, p_j, \theta_{0j}$ are arbitrary constants.

Substituting the function $f$ in  \eqref{e14} and \eqref{e15}-\eqref{e16} into the transformation in  \eqref{eq:trans}, the $N$-solitons of the (2+1)-dimensional time-dependent variable-coefficient KdV system \eqref{eq:vckdv0} can be constructed  explicitly.

\begin{figure}[H]
\centering
\psfig{figure=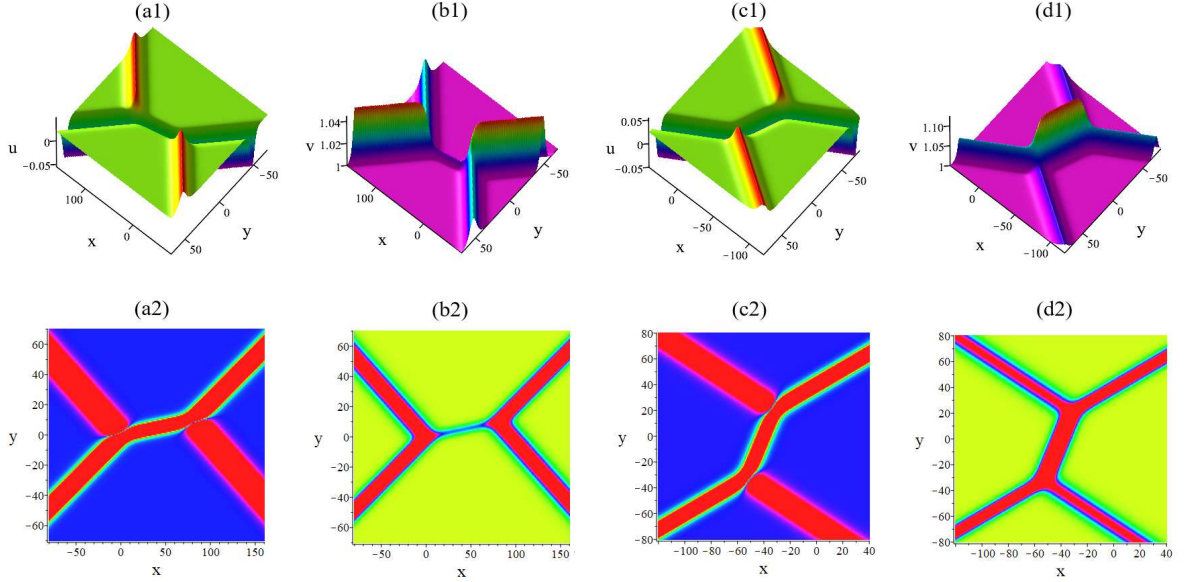, height=8cm,width=16cm}
\caption{ Two-soliton solution given by \eqref{eq:f2} at $t=0$.  Top: 3d plots of $u$ and $v$ versus bottom: corresponding density.  (a1) (a2) (b1) (b2)  H-type interaction with  parameters  $k_1=0.2, k_2=0.3, p_1=0.3, p_2=-0.5, \alpha=0.02, \beta=1$, and hence $e^{A_{12}}=5.6\times10^{-9}$. (c1) (c2) (d1) (d2) X-type interaction with  parameters  $k_1=0.2, k_2=0.3, p_1=0.3, p_2=-0.5, \alpha=0.025, \beta=1$, and hence $e^{A_{12}}=6\times10^{8}$.  }
    \label{fig1}
\end{figure}

{\bf Two-soliton solution.} When $N=2$, Eq. \eqref{e14} reads
\begin{eqnarray}{\label{eq:f2}}
f_2=1+\exp(\theta_1)+\exp(\theta_2)+\exp(\theta_1+\theta_2+A_{12}),
\end{eqnarray}
and the two-solitons of Eq. \eqref{eq:vckdv0} can be obtained via \eqref{eq:trans}. 
This covers the results of \cite{CLP2020} by specifying $\sigma=1, \ \ \mu=1$ and $\rho(t)\equiv 1$.
In the following, we always take $\sigma=1, \ \ \mu=1$ and consider various functions $\rho(t)$.     Fig. \ref{fig1} shows the two-soliton solution with $\rho(t)\equiv 1$. 
 The interacting line solitons form H-type and X-type, which were observed  in ocean waves \cite{MB2012}. Both H-type and X-type
interaction with  long stem, the wave form $u$ have similar structure, while for $v$,  the stem of H-type has a lower amplitude and X-type has the opposite result. 

\begin{figure}[H]
\centering
\psfig{figure=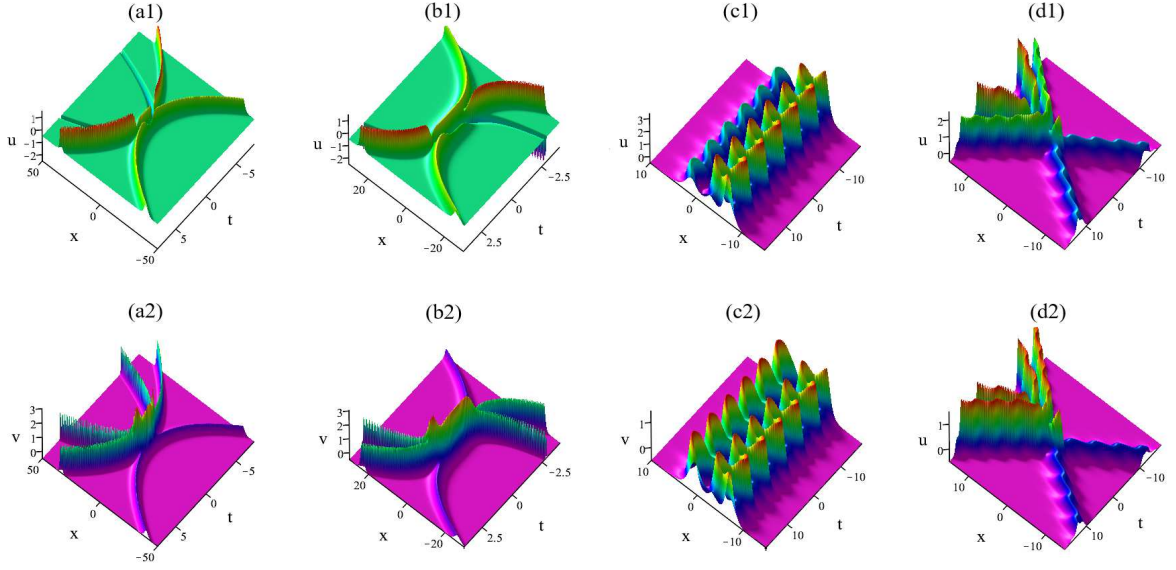, height=8cm,width=16cm}
\caption{ Three-soliton solution given by \eqref{eq:f3} with different variable coefficients $\rho(t)$: (a1) (a2) $\rho(t)=1/t $; (b1) (b2) $\rho(t)=1/{t^2}$; (c1) (c2) $\rho(t)= 1 /{\sin t}$; (d1) (d2) $\rho(t)= 1 /\left({\sin 2t+\tanh t}\right)$. }
    \label{fig2}
\end{figure}

{\bf Three-soliton solution.} When $N=3$, Eq. \eqref{e14} becomes
\begin{eqnarray}{\label{eq:f3}}
\begin{split}
f_3&=1+\exp(\theta_1)+\exp(\theta_2)+\exp(\theta_3)+\exp(\theta_1+\theta_2+A_{12})\\
&\quad+\exp(\theta_1+\theta_3+A_{13}) +\exp(\theta_2+\theta_3+A_{23})+\exp(\theta_1+\theta_2+\theta_3+A_{123}),
\end{split}
\end{eqnarray}
where $A_{123}=A_{12}A_{13}A_{23}$. Substituting $f_3$, i.e., Eq. \eqref{eq:f3}, to \eqref{eq:trans}, three-soliton solution can be obtained. Fig. \ref{fig2} shows novel wave structure with respect to various variable coefficients $\rho(t)$. Top and bottom of the figure illustrate the 3d plots of $u$ and $v$, respectively.  Figs. \ref{fig2} (a1)-(d2) are plotted with the same parameters  $\alpha=-0.5, \beta=-0.5$, $k_1=1, k_2=2, k_3=2, p_1=2, p_2=2$,  but  different $p_3$: (a1) (a2) $p_3=-2$,
(b1) (b2) $p_3=-2$, (c1) (c2) $p_3=4$, (d1) (d2) $p_3=3$. It is observed  that  shapes of waves are influenced by the variable coefficient $\rho(t)$.
\begin{itemize}
\item For $\rho(t)= 1/t$, the three-soliton solution shows the interaction among three parabolic solitons (see Figs. \ref{fig2} (a1) and (a2)).
\item For $\rho(t)= 1 /{t^2}$, the shape of the wave changes from parabolic to cubic, as shown by Figs. \ref{fig2} (b1) and (b2).
\item For trigonometric functions, periodic  waves
are obtained. See Figs. \ref{fig2} (c1), (c2) for $\rho(t)=1/\sin t$ and Figs. \ref{fig2} (d1), (d2) for $\rho(t)=1/\left(\sin 2t +\tanh t\right)$.
\end{itemize}

By specifying the conjugate parameters, two linear solitons can be reduced to one breather. By applying the long wave limit method, two linear solitons can be reduced to one lump solution \cite{LWW2019}. The following theorem can  then be obtained.

\begin{thm}\label{thm:31}
Let
 \begin{align}  {\label{eq:conpar}}
 \nonumber
&p_s=m_s k_s,\quad a_s=l_s \epsilon, \quad s=1,2,\ldots, N,\\
 \nonumber
 &\exp(\theta_{0j})=-1, \quad j=1,2,\ldots, 2M, \\
 &m_n=m_{n+M}^*,\quad n=1,2,\ldots, M,\\
\nonumber
& k_{2M+l}=k_{2M+L+l}^*,\quad l=1,2,\ldots, L,\\
 \nonumber
 &k_{2M+2L+h},\quad h=1,2,\ldots, Q,
  \end{align}
be  constants, where `*' denotes the complex conjugate.  Let $\epsilon\rightarrow 0$,
and the $N$-soliton solution of  Eq. \eqref{eq:vckdv0} can be represented as a combination of  $M$-lump, $L$-breather and $Q$-line solitons, where $N=2M+2L+Q$ and $M, L, Q$ are nonnegative integers \cite{LRW2021}.
\end{thm}

\begin{figure}[H]
\centering
\psfig{figure=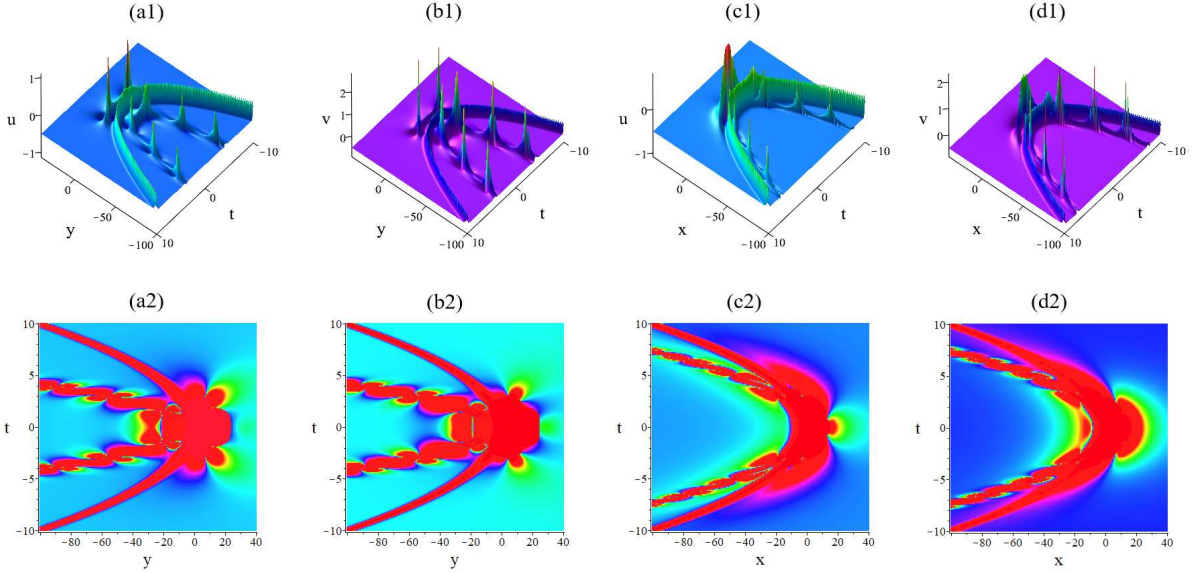, height=8cm, width=16cm}
\caption{ The interaction among one lump, one breather and one line solitons with $M=1, L=1, Q=1$ in Eq. \eqref{eq:conpar}  and with respect to the variable coefficient $\rho(t)= 1 /t$. Top: 3d plots of $u$ and $v$ versus bottom: corresponding density.  The parameters are $k_1=1, k_2=1, k_3=\frac 1 3, k_4=\frac 1 3, k_5=1, p_1=\frac 1 2+i, p_2=\frac 1 2-i, p_3=\frac 1 3+\frac1 3 i, p_4=\frac 1 3-\frac 1 3 i, p_5=1$.}
    \label{fig3}
\end{figure}

Let $M=1, L=1, Q=1$ in \eqref{eq:conpar}, and so $N=5$. Namely, five-soliton solution can be reduced to the interaction among one lump, one breather and one line soliton. 
Figs. \ref{fig3} (a1), (b1) describe the dynamical behavior in the $(y, t)$-
plane when $x=2$, while Figs. \ref{fig3} (c1), (d1) describe the dynamical behavior in the $(x, t)$-
plane when $y=2$.
It is noticed that the lump, breather and line solitons are localized in the parabolic curves and interact with each other.

In the following, we will show several  other special $N$-soliton solutions as corollaries of Theorem \ref{thm:31}.

\begin{cor}
Setting $N=2L$  and defining the parameters
\begin{equation*}
k_j=k_{L+j}^*=\delta_{j}+i\gamma_{j},  p_j=p_{L+j}^*=\kappa_{j}+i \lambda_{j}, \theta_j=\theta_{L+j}^*=\zeta_{j}+i \xi_{j}, \\
\theta_{0j}=(\theta_{0,L+j})^*=\zeta_{0j}+i\xi_{0j},
\end{equation*}
for $ j=1,2,\ldots, L$,
 one can represent the $N$-soliton solution of \eqref{eq:vckdv0} as a combination of $L$-breather solutions.
\end{cor}

\begin{figure}[H]
\centering
\psfig{figure=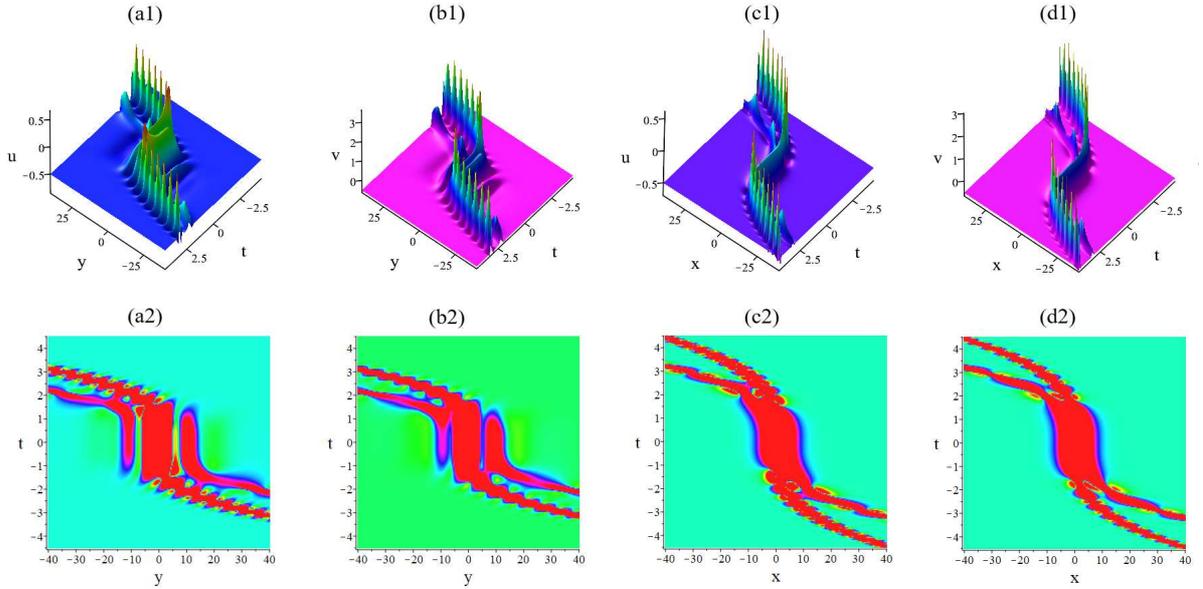, height=8cm,width=16cm}
\caption{  The interaction between two breathers with $L=2$  in \eqref{eq:f2l}  and the variable coefficient $\rho(t)= 1 /{t^2}$. Top: 3d plots of $u$ and $v$ versus bottom: corresponding density.  The parameters are  $k_1=\frac 1 2, k_2=\frac 1 2, k_3=1, k_4=1, p_1=\frac 1 3+\frac 1 3i, p_2=\frac 1 3-\frac 1 3i, p_3=\frac 1 3+\frac 2 3i, p_4=\frac 1 3-\frac 2 3i$.}
    \label{fig4}
\end{figure}

Now, the corresponding representation can be obtained from \eqref{eq:trans}, if we set
\begin{equation}{\label{eq:f2l}}
f_{2L}=\sum_{\varsigma=0,1}\exp\left(\sum_{j=1}^{2L} \varsigma_j \theta_j+\sum_{1\leq s<j}^{2L} \varsigma_s \varsigma_j A_{sj} \right),
\end{equation}
with $\theta_j=\kappa_{j}+i \lambda_{j},$ and $\exp(A_{sj})$ defined by \eqref{e15} and \eqref{e16} respectively,  and consequently
\begin{equation}
\begin{split}
&\zeta_{j}=\delta_{j}x+\kappa_{j} y-\int\frac{\Lambda_j}{\kappa_j^2+\lambda_j^2}d t+\zeta_{0j},\\
&\xi_{j}=\gamma_{j} x+ \sigma_{j} y- \int \frac{\Upsilon_j}{\kappa_j^2+\lambda_j^2}dt+\xi_{0j},
\end{split}
\end{equation}
where $j=1,2,\ldots, L$, and  
\begin{equation*}
\begin{split}
&\Lambda_j=\sigma(\kappa_j^2 + \lambda_j^2)\delta_j^3 + 3\alpha\mu\delta_j^2\kappa_j + 3\Big((\beta\mu-\lambda_j^2\sigma)\kappa_j^2 + (2\alpha\lambda_j \gamma_j + \beta\lambda_j^2)\mu - \sigma\lambda_j^2\gamma_j^2\Big)\delta_j - 3\alpha\mu\lambda_j^2\kappa_j,\\
&\Upsilon_j=-\sigma(\kappa_j^2 + \lambda_j^2)\gamma_j^3 + 3\alpha\mu\gamma_j^2\lambda_j + 3\Big((\delta_j^2\sigma + \beta\mu)\lambda_j^2 + (2\alpha\kappa_j\delta_j + \kappa_j^2\beta)\mu + \sigma\kappa_j^2\delta_j^2\Big)\gamma_j - 3\alpha\mu\delta_j^2\lambda_j.
\end{split}
\end{equation*}
When $L=2$, the four-soliton solution reduces to the interaction between two breathers for $\rho(t)=1/t^2$.
Figs. \ref{fig4} (a1) and (b1) describe the dynamical behavior in the $(y, t)$-
plane when $x=0$, while Figs. \ref{fig4} (c1) and (d1) describe the dynamical behavior in the $(x, t)$-
plane when $y=0$.We notice  that the two breathers have different amplitudes and periods, but both are
with an S-type structure.

\begin{cor}
Set $N=2M$,  and define the parameters
 \begin{align*}
&p_j=k_j m_j,\quad k_j=l_j \epsilon,\quad \theta_{j}^0=i\pi, \quad  j=1,2,\ldots, N,\\
& m_{n}=m_{n+M}^*,\quad n=1,2,\ldots, M.
 \end{align*}
Let $\epsilon\rightarrow 0$, and one can represent  $N$-soliton solution of \eqref{eq:vckdv0} as a combination of $M$-lump solutions \cite{SA1979,ZLT2018}.
\end{cor}

\begin{figure}[H]
\centering
\psfig{figure=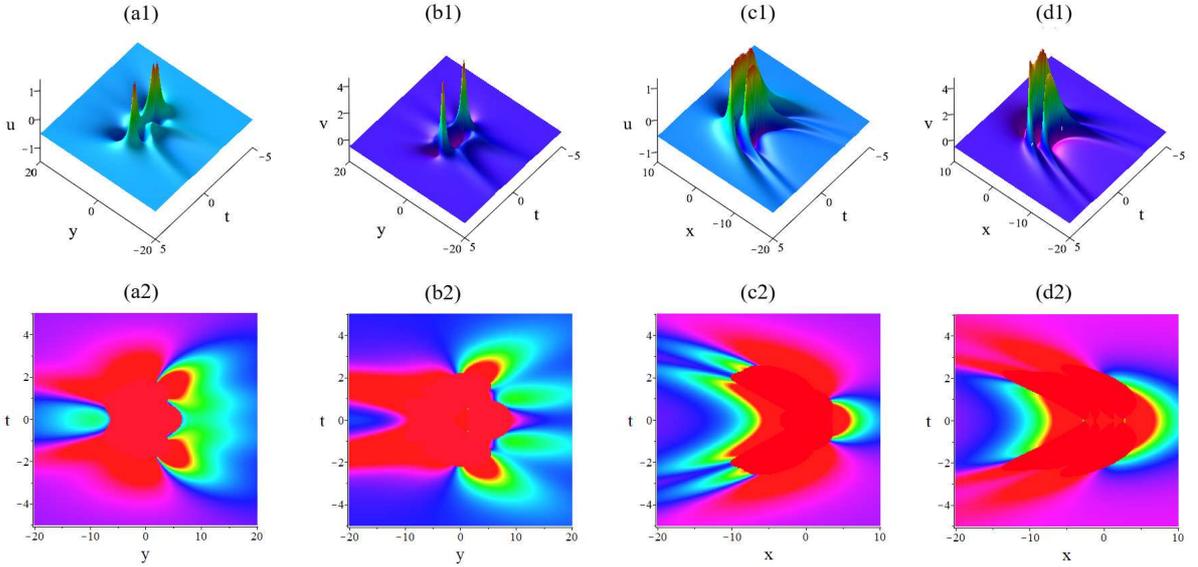, height=8cm,width=16cm}
\caption{ The interaction between two lump solutions  with $M=2$ in \eqref{eq:f2m}  and the variable coefficient $\rho(t)= 1 /{t}$. Top: 3d plots of $u$ and $v$ versus bottom: corresponding density.  The parameters are $k_1=1, k_2=1, k_3=\frac 1 2, k_4=\frac 1 2, p_1=\frac 1 3+\frac 1 4 i, p_2=\frac 1 3-\frac 1 4 i, p_3=\frac 1 4+i, p_4=\frac 1 4-i$. }
    \label{fig5}
\end{figure}

The corresponding representation can be obtained from \eqref{eq:trans} by using
\begin{align}{\label{eq:f2m}}
\begin{split}
f_{2M}&=\prod_{j=1}^{2M}\Theta_j+\frac 1 2\sum_{s,j}^{2M}a_{sj}\prod_{l\neq s,j}^{2M}\Theta_l+\frac 1 {2!2^2}\sum_{s,j,k,m}^{2M}a_{sj}a_{km}\prod_{l\neq s,j,k,m}\Theta_l+\cdots \\
&\quad +\frac 1 {M!2^M}\sum_{s,j,k,m}^{2M}a_{sj}\overbrace{a_{rl}\cdots a_{wn}}^M\prod_{p\neq s,j,r,l,\ldots, w, n}^{2M}\Theta_p+\cdots,
\end{split}
\end{align}
where 
\begin{align} 
\begin{split}
&\Theta_s=x+p_s y-3\mu\int\frac{\beta p_s+\alpha} {\rho(t) p_s} dt,  \quad s=1,2,\ldots, 2M,\\
&a_{sj}=\frac{2\sigma p_s p_j(p_s+p_j)}{\alpha \mu(p_s-p_j)^2}, \quad 1 \leq s< j\leq 2M,
\end{split}
\end{align}
with $p_{s}$ ($s=1,2, \ldots, 2M$)  arbitrary complex constants. Let $M=2$ in \eqref{eq:f2m}. Fig. \ref{fig5} shows that the four-soliton solution reduces to the interaction between two lump solutions.
The effect of the variable coefficient on the interactions is provided  on the $(y, t)$-plane for $x=0$, i.e., (a1), (a2), (b1), (b2), and $(x, t)$-plane for $y=0$, i.e., (c1), (c2), (d1), (d2). Obviously the variable coefficient $\rho(t)$, chosen as $1/t$ in Fig. \ref{fig5},  is closely related to wave shapes.


\begin{rem}
When $N=2M$ is even, the $N$-soliton solution can amount to M-lump or M-breather solitons. When $N=2M+1$ is odd, the hybrid solution has at least one line soliton.
\end{rem}

\section{Symmetry analysis and similarity reduction of the integrable variable-coefficient  KdV system }

\label{sec:sym}

In this section, we conduct symmetry analysis and in particular similarity reduction of the  (2+1)-dimensional  integrable variable-coefficient  KdV system \eqref{eq:vckdv0}.

\subsection{Lie point symmetries}
Consider Lie point symmetries with infinitesimal generators
\begin{eqnarray}
X=\tau (x,y,t,u,v)\partial_t+\xi(x,y,t,u,v)\partial_x+\eta  (x,y,t,u,v)\partial_y+\varphi (x,y,t,u,v)\partial_u+ \psi (x,y,t,u,v)\partial_v
\end{eqnarray}
with coefficients to be determined by the linearized symmetry condition \cite{Olver1993}. Direct but length calculation amounts to
\begin{eqnarray}
\begin{split}
&\tau=f_1(t), \\
&\xi=\left(\frac 1 3 f_1'(t)-\frac13 \frac{\rho'}{\rho}f_1(t)\right)x+f_2(t),\\
&\eta=f_3(y),\\
& \varphi=-\left(\frac 1 3 f_1'(t)-\frac 1 3 \frac {\rho'}{\rho}f_1(t)+f_3'(y)\right)u,\\
&\psi=\frac {1}{9\mu}\left(\rho f_1''(t) x-\rho'' f_1(t) x-\rho' f_1'(t) x+\frac{\rho'^2}{\rho} f_1(t) x-6\mu f_1'(t)v+6\mu \frac {\rho'}{\rho}f_1(t) v+3\rho f_2'(t)\right),
\end{split}
\end{eqnarray}
where $f_1(t)$, $f_2(t)$ and $f_3(y)$ are arbitrary function of $t$ and $y$, respectively. Hence Lie point symmetries of the variable-coefficient KdV system \eqref{eq:vckdv0} are infinite dimensional depending on arbitrary functions, and are spanned by the following infinitesimal generators
\begin{eqnarray}
\begin{split}
&f_1(t)\frac{\partial}{\partial t} + \frac 1 3 \left(f_1'(t)-\frac{\rho'}{\rho}f_1(t)\right)x\frac{\partial}{\partial x}-\frac 1 3 \left( f_1'(t) - \frac{\rho'}{\rho} f_1(t)  \right)u\frac{\partial}{\partial u}\\
&\quad + \frac{1}{9\mu}\left(  \left(\rho^2\left(\frac{f_1'(t)}{\rho}\right)'-\rho\left(\frac{\rho'}{\rho}\right)'f_1(t)\right) x -6\mu\left(f_1'(t)-\frac{\rho'}{\rho}f_1(t)\right) v\right)\frac{\partial}{\partial v},\\
&f_2(t)\frac{\partial}{\partial x}+\frac {1}{3\mu}\rho f_2'(t)\frac{\partial}{\partial v},\\
& f_3(y)\frac{\partial}{\partial y}-f_3'(y)u\frac{\partial}{\partial u}.
\end{split}
\end{eqnarray}
 For simplicity, we will choose linear functions  $f_1(t)=c_1t+c_2$, $f_2(t)=c_3t+c_4$ and $f_3(y)=c_5y+c_6$ in the rest of the paper, where $c_i$, $i=1,2, \ldots, 6$ are  constants.
Consequently, these symmetries of the variable-coefficient KdV system are generated by the following  infinitesimal generators
\begin{eqnarray}\label{infgen6}
\begin{split}
&X_1=\frac{\partial}{\partial x},\quad X_2=\frac{\partial}{\partial y},\quad X_3=t\frac{\partial}{\partial x}+\frac{\rho}{3\mu}\frac{\partial}{\partial v},\quad X_4=y\frac{\partial}{\partial y}-u\frac{\partial}{\partial u},\\
&X_5=\frac{\partial}{\partial t}-\frac13\frac{\rho'}{\rho} x\frac{\partial}{\partial x}+\frac13\frac{\rho'}{\rho}u\frac{\partial}{\partial u}+\frac {1}{9\mu}\left(-\rho''x+\frac{\rho'^2 }{\rho} x+6\mu \frac{\rho'}{\rho} v\right)\frac{\partial}{\partial v},\\
&X_6=t\frac{\partial}{\partial t}+\frac 1 3\left(1-\frac{\rho' }{\rho}t\right)x\frac{\partial}{\partial x}-\frac 1 3 \left(1-\frac{\rho' }{\rho}t\right)u\frac{\partial}{\partial u}-\frac {1}{9\mu}\left(\rho''tx+\rho'x-\frac{\rho'^2}{\rho}tx+6\mu v-6\mu  \frac{\rho'}{\rho}tv \right)\frac{\partial}{\partial v}.
\end{split}
\end{eqnarray}

\subsection{Similarity reductions}
In this subsection, we will study similarity reductions of  the variable-coefficient KdV system \eqref{eq:vckdv0} by using each of the symmetries \eqref{infgen6}. Certainly their linear combinations may lead to further interesting solutions.

(1) $X_1=\frac {\partial}{\partial x}$.  The corresponding invariants are  $t,y,u,v,$,
and the group-invariant solution is
\begin{eqnarray}
u=U(y),\ \ v=V(t),
\end{eqnarray}
where $U(y)$ and $V(t)$ are arbitrary function about $y$ and $t$, respectively.

(2) $X_2=\frac {\partial}{\partial y}$. Direct calculation gives the following group-invariant solution
\begin{eqnarray}
u=U(t),\ \ v=-\frac{1}{3\mu} \frac{ U'(t) }{ U(t)}\rho x,
\end{eqnarray}
where $U(t)$ is an arbitrary function.

\begin{figure}[H]
\centering
\psfig{figure=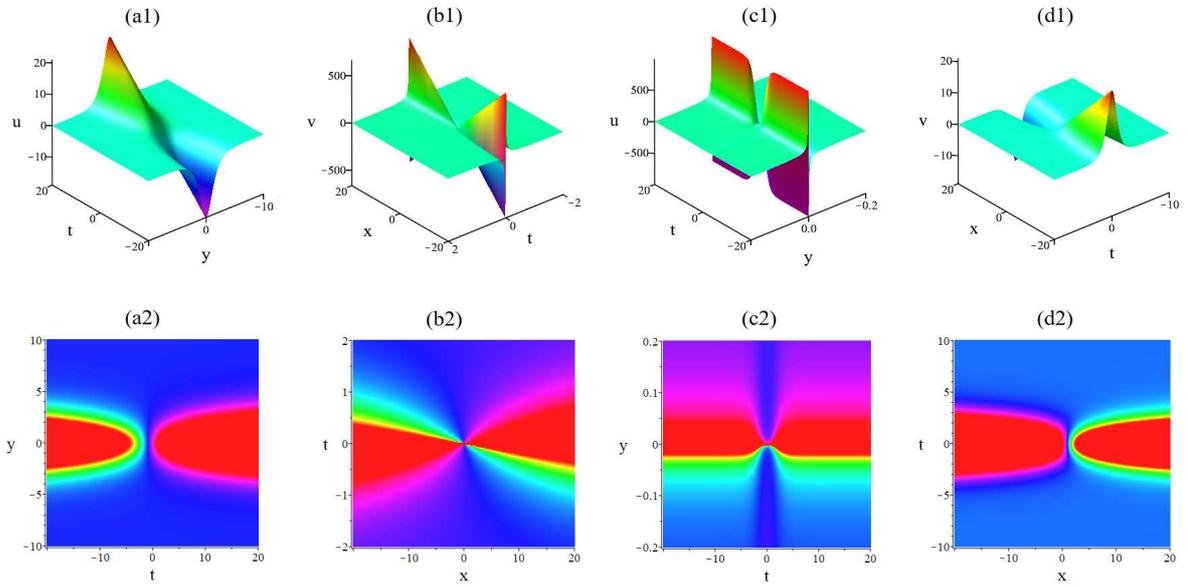, height=8cm,width=16cm}
\caption{  The interaction between two soliton solutions given by \eqref{sol:x3} (figures (a1), (a2), (b1), (b2)) and \eqref{sol:x4} (figures (c1), (c2), (d1), (d2)). Top: 3d plots of $u$ and $v$ versus bottom: corresponding density. }
    \label{fig6}
\end{figure}

(3) $X_3=t\frac{\partial}{\partial x}+\frac{\rho}{3\mu}\frac{\partial}{\partial v}$.
The characteristic equation for determining the invariants reads
\begin{eqnarray}
\frac {dx}{t}=\frac {dv}{{\rho}/{3\mu}},
\end{eqnarray}
amounting to the following invariants
\begin{equation}\label{eq:x3}
t,\quad y,\quad U(y,t)=u,\quad V(y,t)=\frac{3\mu}{\rho} \left(v-\frac{1}{3\mu } \frac{\rho}{t}x\right).
\end{equation}
Substituting \eqref{eq:x3} back to the system \eqref{eq:vckdv0} gives
\begin{eqnarray}{\label{sol:x3}}
u=R(y)t+K(y),\ \ v=\frac{1}{3\mu } \frac{\rho}{ t}x+\frac{1}{3\mu} \rho V(t),
\end{eqnarray}
where $R(y)$, $K(y)$ and $V(t)$ are arbitrary function about $y$ and $t$, respectively. To illustrate this solution, we choose $R(y)$=sech$(y)$, $K(y)$=sech$(y)$, $V(t)$=sech$(t)$, $\rho$=sech$(t)$ and $\mu=1$. The figures of $u$ and $v$ are shown in Figs. \ref{fig6} (a1), (a2),(b1) and (b2). The interaction between two soliton, with the opposite amplitude in the $y$-direction, and with the same amplitude in the $x$-direction, can be observed.

(4) $X_4=y\frac{\partial}{\partial y}-u\frac{\partial}{\partial u}$. 
Similar computation give the following group-invariant solution
\begin{eqnarray}{\label{sol:x4}}
u=cy^{-1}e^{-3\mu U(t)},\ \ v=U'(t)\rho x+V(t),
\end{eqnarray}
where $U(t)$ and $V(t)$ are arbitrary function of $t$, and $c$ is a constant. We choose $R(t)$=sech$(t)$, $V(t)$=sech$(t)$, $c=1$, $\rho=1$ and $\mu=1$, and the resulting solution is shown in Figs. \ref{fig6} (c1), (c2), (d1) and (d2).  The interaction between two solitons, with the same amplitude in the $y$-direction, and with the opposite amplitude in  the $x$-direction can be noticed.

(5) $X_5=\frac{\partial}{\partial t}-\frac13\frac{\rho'}{\rho}x \frac{\partial}{\partial x}+\frac13\frac{\rho'}{\rho}u\frac{\partial}{\partial u}+\frac {1}{9\mu}\left(-\rho''x+\frac{\rho'^2}{\rho}x+6\mu\frac{\rho'}{\rho}v\right)\frac{\partial}{\partial v}$. Consider the characteristic equations
\begin{eqnarray}
\frac {dx}{-\frac13\frac {\rho'}{\rho}x}=\frac {dt}{1}=\frac {du}{\frac13\frac {\rho'}{\rho}u}=\frac {dv}{\frac 1 {9\mu}\left(-\rho''x+\frac {\rho'^2}{\rho}x+6\mu\frac {\rho'}{\rho}v\right)},
\end{eqnarray}
whose solution give the invariants
\begin{eqnarray}
\rho^{\frac 1 3}x, \  \ y, \ \ xu,\ \ \frac {1} {9\mu}\frac{\rho'}{\rho} \rho^{\frac 1 3}x  +\rho^{-\frac 2 3}v.
\end{eqnarray}
To conduct the reduction, we choose
\begin{eqnarray}{\label{inv:x5}}
 r=\rho^{\frac 1 3}x, \ \ y=y,\quad U=\frac { x u} {r},\ \ V=\frac {1} {9\mu}\frac{\rho'}{\rho}r+\rho^{-\frac 2 3}v,
\end{eqnarray}
which are substituted back to \eqref{eq:vckdv0}, yielding
\begin{equation}\label{eq:redx5}
 \left\{
\begin{array}{l}
3\mu \left(UV\right)_r+\sigma U_{rrr}=0, \vspace{0.2cm}\\
U_r-V_y=0.\\
\end{array}
\right.
\end{equation}
The reduced system \eqref{eq:redx5} is still difficult to be solved immediately. In the following, we will conduct one more step of reduction. 
As shown in (i) below that the system \eqref{eq:redx5} can be reduced to one PDE, and we will conduct the symmetry reduction for \eqref{eq:redx5} and the PDE, i.e., \eqref{eq:newintpde}, separately. They are related as local and nonlocal symmetries of differential equations (see, e.g., \cite{BTS2005} and references therein).

(i) {\bf Symmetries of the PDE \eqref{eq:newintpde}.} The first equation of \eqref{eq:redx5} can be integrated with respect to $r$, yielding
\begin{eqnarray}{\label{eq:redx5-1}}
V=\frac{a(y)-\sigma U_{rr}}{3\mu U},
\end{eqnarray}
which is then substituted to the second equation. Consequently, the system \eqref{eq:redx5} turns into a single PDE
\begin{eqnarray}\label{eq:newintpde}
U_r=\left(\frac{a(y)-\sigma U_{rr}}{3\mu U}\right)_y,
\end{eqnarray}
where $a(y)$ is an arbitrary function.

\begin{rem}
The  PDE \eqref{eq:newintpde} seems new to us and is potentially integrable by the ARS conjecture \cite{ARS1980}. Indeed, it passes the Painlev\'e test.
\end{rem}
Expanding the derivatives, Eq. \eqref{eq:newintpde} reads
%
\begin{eqnarray}{\label{eq:redx5-2}}
a'(y)U-a(y)U_y-3 \mu U^2 U_r +\sigma U_yU_{rr} -\sigma UU_{rry}=0.
\end{eqnarray}
Lie point symmetries of  \eqref{eq:redx5-2} are generated by the infinitesimal generators
\begin{eqnarray}\label{symv123}
\begin{split}
&Y_1=\frac {\partial} {\partial r},\quad Y_2= \frac 1 {a(y)}\frac {\partial} {\partial y}+\frac{a'(y)}{a^2(y)}U \frac {\partial} {\partial U},\\
&Y_3=r\frac {\partial} {\partial r}-\frac 3{a(y)}\left(\int a(y)dy\right)\frac {\partial} {\partial y}+U\left(2-\frac{3a'(y)}{a^2(y)}\int a(y)dy\right)\frac {\partial} {\partial U},
\end{split}
\end{eqnarray}
where we assume $a(y)\neq 0$.  

(i-1) We firstly use $Y_1+c_0 Y_2$ to reduce Eq. \eqref{eq:redx5-2}, where $c_0$ is constant. The invariants are $z={c_0} r-\int a(y) dy$ and $R(z)=U(r,y)/a(y)$, and the  Eq. \eqref{eq:redx5-2} is reduced to the third-order ODE
\begin{eqnarray}
 -\sigma c_0^2 R'''R+c_0^2 R''R'+3c_0 \mu R' R^2 -R'=0.
\end{eqnarray}
Dividing by $R^2$ on both sides and integrating it with respect to $z$, we obtain a second-order ODE
\begin{eqnarray}\label{eq:v122ndode}
 -\sigma c_0^2 R''+3c_0 \mu R^2+c_1 R+1=0,
\end{eqnarray}
By a translation $\widetilde{R}={R}+c_2$, Eq. \eqref{eq:v122ndode} becomes
\begin{eqnarray}\label{twoorder}
 -\sigma c_0^2 \widetilde{R}''+3c_0 \mu \widetilde{R}^2+(c_1-6\mu c_0 c_2) \widetilde{R}+3\mu c_0c_2^2-c_1c_2+1=0.
\end{eqnarray}
It is equivalent to the reduced ODE of the KdV equation for searching its traveling wave solutions (see Example 3.4 of \cite{Olver1993}), that will appear later, i.e., Eq. \eqref{eq:reduv1}, followed by some special  solutions.

(i-2) The invariants with respect to $Y_3$ are
\begin{equation}
z=r\left(\int a(y)dy\right)^{\frac 1 3},\quad R(z)=\frac{1}{a(y)}\left(\int a(y)dy\right)^{\frac 2 3}U(r,y).
\end{equation}
Now the  Eq. \eqref{eq:redx5-2} is reduced to a third-order ODE
\begin{eqnarray}\label{eq:v3-3rdode}
\begin{split}
 -\sigma zRR'''+ \sigma zR'R''-2\sigma RR''-9\mu R^2R' - zR'+2 R=0.
 \end{split}
\end{eqnarray}
By the transformation
\begin{equation}
R=(zW)',
\end{equation}
 Eq. \eqref{eq:v3-3rdode} becomes
\begin{equation}\label{intode}
\sigma W'''=-3\sigma \frac{W''}{z}-9\mu (W')^2-9\mu \frac{WW'}{z}+c_1\frac{W'}{z^2}+c_1\frac{W}{z^3}+\frac{1}{z},
\end{equation}
where $c_1$ is a constant of integration. Eq. \eqref{intode} is integrable as passing the Painlev\'e test and is in the form of  Chazy's classification on third-order Painlev\'e equations
of the polynomial type (see, e.g., Eq. (2.1) of  \cite{Cos2000}). However, some of the coefficients are locally analytic except $z= 0$. Furthermore, Eq.  \eqref{intode}  is non-autonomous and seems not included in the 13 classes  introduced by Chazy in  \cite{Cha1911}.

\begin{rem} If $a(y)=0$, singularity appears in the symmetries \eqref{symv123}. Now Eq. \eqref{eq:redx5-2} becomes
\begin{eqnarray}{\label{eq:redx5-3}}
-3 \mu U^2 U_r +\sigma U_y U_{rr} -\sigma UU_{rry}=0,
\end{eqnarray}
and its Lie point symmetries are generated by the infinitesimal generators
\begin{eqnarray}{\label{symv123-3}}
\begin{split}
Y_1= \frac {\partial} {\partial r},\quad  Y_2=g(y)\frac {\partial} {\partial y}- U g'(y) \frac {\partial} {\partial U},\quad Y_3=r\frac {\partial} {\partial r}-U\frac {\partial} {\partial U}.
\end{split}
\end{eqnarray}
For simplicity, let $g(y)=1$, and consider $Y_1+c_0Y_2$ that is associated to traveling wave solutions.
Choose the invariants as
\begin{equation}
z=c_0r-y,\quad U.
\end{equation}
The Eq. \eqref{eq:redx5-3} is reduced to
\begin{equation}
\sigma c_0U''-3\mu U^2+c_1U=0,
\end{equation}
where $U=U(z)$ and $c_1$ is a constant. This ODE is equivalent to \eqref{eq:v122ndode} by a translational transformation of $U$.

In addition, we take $g(y)=y$, and consider  $Y_2+cY_3$, which corresponding to the scale invariance. The invariants are $z=r y^{-c}$ and $R(z)=y^{c+1}U(r,y)$, and the  Eq. \eqref{eq:redx5-3} is reduced to the third-order ODE
\begin{eqnarray}\label{eq:intode2}
\begin{aligned}
-\sigma c z R R''' +\sigma c z R'R''-2\sigma c R R'' +3\mu R^2R'=0.
\end{aligned}
\end{eqnarray}
Similar to the derivation of Eq. \eqref{intode}, introducing $R=(zW)'$, Eq. \eqref{eq:intode2} is equivalent to
\begin{equation}\label{ode:p3-2}
\begin{aligned}
\sigma c W'''=-3\sigma c\frac{W''}{z} +3\mu(W')^2+3\mu\frac{WW'}{z}+c_1\frac{W'}{z^2}+c_1\frac{W}{z^3},
\end{aligned}
\end{equation}
where $c_1$ is a constant of integration. It is again a third-order Painlev\'e equation  of polynomial type, which passes the Painlev\'e test. It is equivalent to Eq. \eqref{intode} except  a $1/z$ term.


\end{rem}


(ii) {\bf Symmetries of the potential system \eqref{eq:redx5}.} Its Lie point symmetries are generated by
\begin{eqnarray}
Y_1=\frac {\partial} {\partial r}, \ \  Y_2=h(y) \frac {\partial} {\partial y}-h'(y) U \frac {\partial} {\partial U},\quad Y_3=r\frac {\partial} {\partial r}-U\frac {\partial} {\partial U}- 2V\frac {\partial} {\partial V},
\end{eqnarray}
where $h(y)$ is an arbitrary function.

(ii-1) Consider a special case  by taking $h(y)=1$, and the second generator  becomes
\begin{eqnarray}
Y_2=\frac {\partial} {\partial y}.
\end{eqnarray}
Consider reductions related to $Y_1+c_0 Y_2$, i.e., traveling wave type of solutions, where $c_0$ is a constant. The invariants are $z=c_0 r-y$, and $U$,  $V$. Now the potential system \eqref{eq:redx5} becomes
\begin{equation}\label{eq:reduv}
 \left\{
\begin{array}{l}
3\mu  U' V+3\mu  U V'+\sigma c_0^2 U'''=0, \vspace{0.2cm}\\
c_0U'+V'=0.\\
\end{array}
\right.
\end{equation}
Both equations in \eqref{eq:reduv} can be integrated once and the system is equivalent to the following second-order ODE
\begin{eqnarray}\label{eq:reduv1}
-\sigma c_0^2U''(z)+3\mu c_0 U^2-3 \mu c_1 U -c_2 =0.
\end{eqnarray}
Similar to Eq. \eqref{eq:v122ndode}, Eq. \eqref{eq:reduv1} can also be integrated once, amounting to
\begin{eqnarray}\label{firint}
 \frac 1 2\sigma c_0^2U'^2(z)=\mu c_0 U^3-\frac 3 2 \mu c_1 U^2 -c_2 U+c_3,
\end{eqnarray}
where $c_1,c_2,c_3$ are integration constants.  If $c_2\neq 0$, Eq. \eqref{eq:reduv1} is equivalent to
 Eq. \eqref{twoorder}. 
%
In the following, we list its special solutions by properly assigning the parameters.
\begin{itemize}
\item Assume $\mu=-1$ without loss of generality, and further let $c_2=0$ and $c_3=0$. A special solution of Eq. \eqref{firint} is given by
\begin{eqnarray}
 U=\frac{3c_1}{2c_0}\mathrm{sech}^2\left(\frac{\sqrt{3\sigma c_1}}{2\sigma c_0}(z+\delta)\right),
\end{eqnarray}
where $\delta$ is a phase shift and we assumed $\sigma c_1$ to be positive. This leads to solutions of the system \eqref{eq:vckdv0} as
\begin{eqnarray}\label{sol:v5ii-1}
\begin{split}
u(x,y,t)&=\frac{3c_1}{2c_0} \rho^{\frac 1 3}\mathrm{sech}^2\left(\frac{\sqrt{3\sigma c_1}}{2\sigma c_0}\left(c_0 \rho^{\frac 1 3} x-y+\delta\right)\right),\\
 v(x,y,t)&=-\frac{3c_1}{2}\rho^{\frac 2 3}\mathrm{sech}^2\left(\frac{\sqrt{3\sigma c_1}}{2\sigma c_0}\left(c_0 \rho^{\frac 1 3} x-y+\delta\right)\right)+c_4 \rho^{\frac 2 3}-\frac{1}{9\mu}\rho' x.
\end{split}
\end{eqnarray}
where $c_4$ is an integration constant.
\item If we only assume $U(z)$ to be bounded similar as \cite{Olver1993}, Eq. \eqref{firint} has the periodic cnoidal wave solutions
\begin{eqnarray}\label{cnsolution}
 U=b_2+(b_3-b_2)\mathrm{cn}^2\left(\sqrt{-\frac{\mu(b_3-b_1)}{2\sigma c_0}}(z+\delta),k \right),
\end{eqnarray}
where cn is the Jacobi elliptic function of modulus $k=\sqrt{\frac{b_3-b_2}{b_3-b_1}}$, and $b_1<b_2<b_3$ are the roots of the cubic polynomial on the right-hand side of Eq. \eqref{firint}. The corresponding periodic wave solutions of the system \eqref{eq:vckdv0} are
\begin{eqnarray}\label{cnsol}
\begin{split}
u(x,y,t)&=\rho^{\frac 1 3}\left(b_2+(b_3-b_2)\mathrm{cn}^2\left(\sqrt{-\frac{\mu(b_3-b_1)}{2\sigma c_0}}\left(c_0 \rho^{\frac 1 3}x-y+\delta\right),k \right)\right),\\
 v(x,y,t)&=-c_0\rho^{\frac 2 3}\left(b_2+(b_3-b_2)\mathrm{cn}^2\left(\sqrt{-\frac{\mu(b_3-b_1)}{2\sigma c_0}}\left(c_0 \rho^{\frac 1 3}x-y+\delta\right),k \right)\right)\\
 &\quad +c_4 \rho^{\frac 2 3}-\frac{1}{9\mu}\rho'x.
\end{split}
\end{eqnarray}
Here, we assumed $\mu\sigma c_0$ to be negative.
\end{itemize}

\begin{figure}[H]
\centering
\psfig{figure=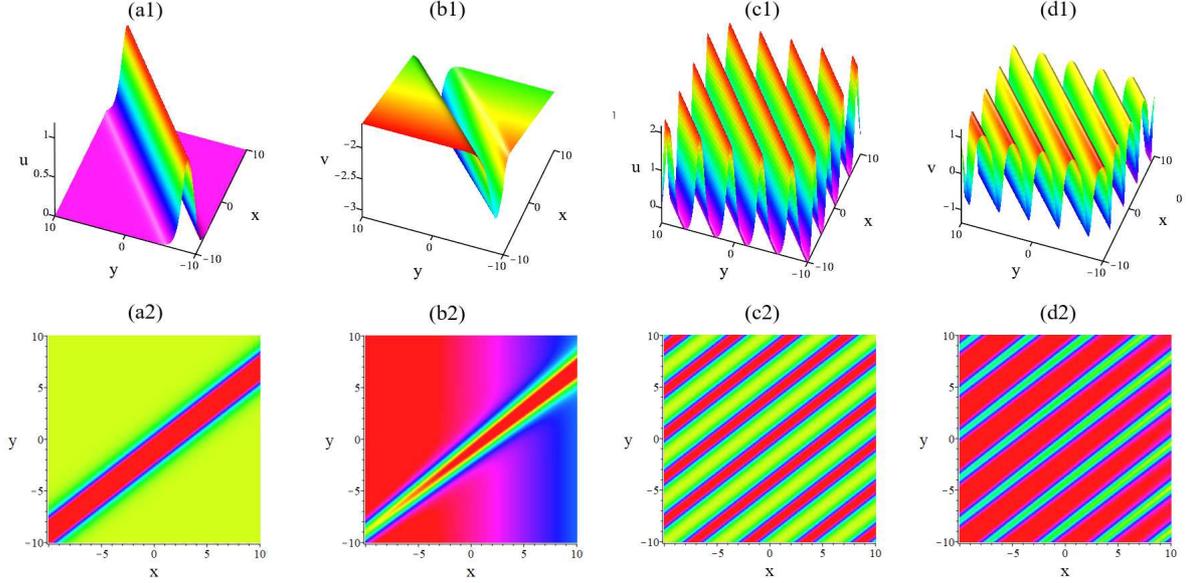, height=8cm,width=16cm}
\caption{  The traveling wave  solutions \eqref{sol:v5ii-1} (figures (a1), (a2), (b1), (b2)) and periodic wave solutions \eqref{cnsol} (figures (c1), (c2), (d1), (d2)). Top: 3d plots of $u$ and $v$ versus bottom: corresponding density. }
    \label{TCsol}
\end{figure}

The traveling wave solutions \eqref{sol:v5ii-1} and periodic wave solutions \eqref{cnsol} of the system \eqref{eq:vckdv0}   are shown in Fig. \ref{TCsol} at the time $t=2$ and with variable coefficient $\rho=1/t$. In Figs. \ref{TCsol} (a1), (a2), (b1) and (b2), the parameters are $\delta=\sigma=c_0=c_1=1, c_4=-3$, while in (c1), (c2), (d1) and (d2), the parameters are $\delta=0, \sigma=c_0=c_1=c_3=c_4=1, c_2=-3, \mu=-1$.

(ii-2) If we choose  $h(y)=y$, then $Y_2+cY_3$ corresponds to the scale invariance, the invariants of which are $z=ry^{-c}$, $R(z)=y^{c+1}U(r,y)$, $F(z)=r^2V(r,y)$. Then, Eq. \eqref{eq:redx5} is reduced to the third-order ODE
\begin{eqnarray}\label{eq:v23-3rdode}
\begin{split}
 \sigma zRR'''-\sigma zR'R''+2\sigma RR''-\frac{3\mu} c R^2R' - c_0zR'+2c_0 R=0.
 \end{split}
\end{eqnarray}
 It is equivalent to Eq. \eqref{eq:v3-3rdode} by a scaling of $R$ providing the constant of integration  $c_0\neq 0$. If $c_0=0$, defining
 \begin{equation}
 R=(zW)',
 \end{equation}
Eq. \eqref{eq:v23-3rdode} becomes
  \begin{equation}\label{ode:p3-3}
\sigma W'''=-3\sigma \frac{W''}{z}+\frac{3\mu}{c} (W')^2+\frac{3\mu}{c}\frac{WW'}{z}+c_2\frac{W'}{z^2}+c_2\frac{W}{z^3}
\end{equation}
with $c_2$ a constant of integration. It is exactly Eq. \eqref{ode:p3-2}.

(6) $X_6=t\frac{\partial}{\partial t}+\frac 1 3\left(1-\frac{\rho' }{\rho}t\right)x\frac{\partial}{\partial x}-\frac 1 3 \left(1-\frac{\rho' }{\rho}t\right)u\frac{\partial}{\partial u}-\frac {1}{9\mu}\left(\rho''tx+\rho'x-\frac{\rho'^2}{\rho}tx+6\mu v-6\mu \frac{\rho'}{\rho} tv \right)\frac{\partial}{\partial v}$. The characteristic equations are
\begin{eqnarray} 
\frac {dx}{\frac x 3\left(1-\frac{\rho' t}{\rho}\right)}=\frac {dt}{t}=\frac {du}{\frac u 3 \left(\frac{\rho' t}{\rho}-1\right)}=\frac {dv}{-\frac {1}{9\mu}\left(\rho''tx+\rho'x-\frac{\rho'^2}{\rho}tx+6\mu v-6\mu  \frac{\rho'}{\rho}tv \right)},
\end{eqnarray}
solving that gives the invariants
\begin{eqnarray}
\left(\frac{\rho}{t}\right)^{\frac 1 3}x, \ \ y, \ \  x u,\ \ \frac {1} {9\mu} \frac{\rho'}{\rho} \left(\frac{\rho}{t}\right)^{\frac 1 3}tx+\left(\frac{\rho}{t}\right)^{-\frac 2 3}v.
\end{eqnarray}
We choose the invariant variables as
\begin{eqnarray}{\label{v6inv}}
r=\left(\frac{\rho}{t}\right)^{\frac 1 3}x, \quad y=y,\ \  U=\frac {xu} r,\ \ V=\frac {1} {9\mu} \frac{\rho'}{\rho} \left(\frac{\rho}{t}\right)^{\frac 1 3}tx+\left(\frac{\rho}{t}\right)^{-\frac 2 3}v,
\end{eqnarray}
which are substituted into \eqref{eq:vckdv0}, yielding
\begin{equation}\label{eq:v6}
 \left\{
\begin{array}{l}
-\frac 1 3 (U+ U_r r)+3\mu( U V)_r+\sigma U_{rrr}=0, \vspace{0.2cm}\\
U_r-V_y=0.\\
\end{array}
\right.
\end{equation}

\begin{rem}
The system \eqref{eq:v6} is equivalent to \eqref{eq:redx5} by the transformation
\begin{equation}
U\mapsto U,\quad V\mapsto V+\frac{r}{9\mu}.
\end{equation}
In other words, if $(U,V)$ is a  solution of \eqref{eq:redx5}, then $(U,V+\frac{r}{9\mu})$  is a solution of \eqref{eq:v6}.   For instance, from \eqref{sol:v5ii-1}, we  obtain other solutions of \eqref{eq:vckdv0} as follows
\begin{eqnarray} 
\begin{split}
u(x,y,t)&=\frac{3c_1}{2c_0} \left(\frac{\rho}{t}\right)^{\frac 1 3}\mathrm{sech}^2\left(\frac{\sqrt{3\sigma c_1}}{2\sigma c_0}\left(c_0\left(\frac{\rho}{t}\right)^{\frac 1 3}x-y+\delta\right)\right),\\
 v(x,y,t)&=-\frac{3c_1}{2}\left(\frac{\rho}{t}\right)^{\frac 2 3}\mathrm{sech}^2\left(\frac{\sqrt{3\sigma c_1}}{2\sigma c_0}\left(c_0\left(\frac{\rho}{t}\right)^{\frac 1 3}x-y+\delta\right)\right)+c_4 \left(\frac{\rho}{t}\right)^{\frac 2 3}-\frac{1}{9\mu}\rho'x+\frac{1}{9\mu}\frac{\rho}{t}x,
\end{split}
\end{eqnarray}
and corresponding to  \eqref{cnsol}, we obtain
\begin{eqnarray} 
\begin{split}
u(x,y,t)&=\left(\frac{\rho}{t}\right)^{\frac 1 3}\left(b_2+(b_3-b_2)\mathrm{cn}^2\left(\sqrt{-\frac{\mu(b_3-b_1)}{2\sigma c_0}}\left(c_0\left(\frac{\rho}{t}\right)^{\frac 1 3}x-y+\delta\right),k \right)\right),\\
 v(x,y,t)&=-c_0\left(\frac{\rho}{t}\right)^{\frac 2 3}\left(b_2+(b_3-b_2)\mathrm{cn}^2\left(\sqrt{-\frac{\mu(b_3-b_1)}{2\sigma c_0}}\left(c_0\left(\frac{\rho}{t}\right)^{\frac 1 3}x-y+\delta\right),k \right)\right)\\
 &\quad +c_4 \left(\frac{\rho}{t}\right)^{\frac 2 3}-\frac{1}{9\mu}\rho'x+\frac{1}{9\mu}\frac{\rho}{t}x.
\end{split}
\end{eqnarray}

\end{rem}


\section{Conclusions}


\label{sec:con}

In this paper,  a (2+1)-dimensional integrable KdV system with time-dependent variable coefficient was studied. Its integrability is analyzed by Painlev\'e analysis. $N$-soliton solutions of the (2+1)-dimensional variable-coefficient KdV system were obtained by using Hirota's bilinear method. In particular, by choosing appropriate parameters on the $N$-soliton solutions, novel wave interaction phenomena were discovered, e.g., the soliton solutions shown in Figs. \ref{fig1}-\ref{fig2}, the hybrid interaction of line, lump and breather solitons illustrated by Fig. \ref{fig3}, the interaction of two breathers (Fig. \ref{fig4}), and the interaction of  two lump solutions (Fig. \ref{fig5}). Furthermore, group-invariant solutions are derived by similarity reduction,  for instance, an interaction between two solitons in Fig. \ref{fig6} beside other interesting analytic solutions. These results show interesting novel physical features, which should provide new knowledge in the study of  variable-coefficient nonlinear systems.  As a final remark,  the  (2+1)-dimensional integrable variable-coefficient KdV system and the reduced PDE \eqref{eq:newintpde} are among the few examples that can be reduced to third-order Painlev\'e equations.

\section*{Acknowledgements}
L. Peng is grateful to Saburo Kakei, Frank Nijhoff and Ralph Willox for helpful discussions.
Y. Liu was partially supported by the National Natural Science Foundation of China (No. 11905013),  the Beijing Natural Science Foundation (No. 1222005), and Qin Xin Talents Cultivation Program of Beijing Information Science and Technology University (QXTCP C202118). L. Peng was partially supported by JSPS Kakenhi Grant Number JP20K14365,
JST-CREST Grant Number JPMJCR1914, and Keio Gijuku Fukuzawa Memorial Fund.



 \end{document}